\documentclass[10pt,conference]{IEEEtran}

\usepackage{graphicx}

\newcommand{\df}{\stackrel{\bigtriangleup}{=}}

\newtheorem{theorem}{Theorem}

\newtheorem{remark}{Remark}

\begin{document}

\title{On Multiple Hypothesis Testing with  \\ Rejection Option}

\author{
\IEEEauthorblockN{Naira Grigoryan}
\IEEEauthorblockA{IIAP NAS RA \\
Email: nar.gri@gmail.com} \and \IEEEauthorblockN{Ashot Harutyunyan}\IEEEauthorblockA{IIAP NAS RA \\
Email: ashot@iem.uni-due.de} \and \IEEEauthorblockN{Svyatoslav
Voloshynovskiy}
\IEEEauthorblockA{University of Geneva \\
Email: svolos@unige.ch} \and \IEEEauthorblockN{Oleksiy Koval}
\IEEEauthorblockA{University of Geneva \\
Email: oleksiy.koval@unige.ch} }

\maketitle

\begin{abstract}
We study the problem of multiple hypothesis testing (HT) in
view of a rejection option. That model of HT has many different applications.
Errors in testing of $M$ hypotheses regarding the source
distribution with an option of rejecting all those hypotheses are
considered. The source is discrete and arbitra\-ri\-ly varying
(AVS). The tradeoffs among error probability
exponents/re\-lia\-bi\-li\-ties associated with {\it false acceptance of rejection decision} and {\it false rejection of true distribution} are investigated and the optimal
decision strategies are outlined. The main result is
specialized for discrete memoryless sources (DMS) and studied
further. An interesting insight that the analysis implies is the phenomenon (comprehensible in terms of supervised/unsupervised learning) that in optimal discrimination within $M$ hypothetical distributions one permits always lower error than in deciding to decline the set of hypotheses.
Geo\-met\-ric interpretations of the optimal decision
schemes are given for the current and known bounds in multi-HT for
AVS's.

\end{abstract}

\section{Introduction}
Recent impetuous progress in computer and public network infrastructure
as well as in multimedia data manipulating software created an
unprecedented yet often uncontrolled possibilities for multimedia
content modification and redistribution over various public services and
networks including Flickr and YouTube. Since in multiple cases these
actions concern privacy sensitive data, a significant research effort
was made targeting efficient means of their identification as well as
related performance analysis \cite{Haitsma2001}, \cite{Lefebvre2002}, \cite{Koval2007}. While early reported
results \cite{Willems2003} were mostly dedicated to the capacity analysis of
identification systems, more recent considerations are based on multiple HT
framework with a rejection option. Possible examples for binary data statistics are presented in \cite{Varna2008} and \cite{slava-oleksiy1}.
Motivated by the prior art, we extend the problem of content
identification as multiple HT with rejection to a broader
class of source priors including AVS's. Our
analysis lies within the frames of the works by Hoeffding
\cite{HF65}, Csisz\'ar and Longo \cite{CL71}, Blahut \cite{BL74},
Haroutunian \cite {H90}, Birg\'e \cite{B81}, Fu and Shen
\cite{FSH98}, Tuncel \cite{T05}, Grigoryan and Harutyunyan
\cite{GH10} with the aim of specifying the asymptotic
bounds for error probabilities. Those papers do not treat an
option of rejection. In particular, \cite{BL74} characterizes the
optimum relation between two error exponents in binary HT and
\cite{H90} (see also \cite{AH06}, \cite{HHH}) and \cite{T05} study the
multiple ($M>2$) HT for DMS's in terms of logarithmically
asympto\-ti\-cally optimality (LAO) and errors exponents
achievability, respectively. Later advances in the binary and $M$-ary HT for a more general class of sources --
AVS's (see also its coding framework \cite{HarVinck2006}), are the subjects of \cite{FSH98} and \cite{GH10},
respectively. The latter derives also Chernoff bounds for HT
on AVS's and extends the finding by Leang and Johnson \cite{LJ97}
for DMS's. Our work is a further extension of $M$-ary HT for
discrete sources in terms of errors occurring with respect to an additional rejection decision. The focus is on the attainable region of error exponents which tradeoff between the {\it false acceptance of rejection decision} and {\it false rejection of true distribution}. A similar
model of HT with empirically observed statistics for Markov sources
has been explored by Gutman in \cite{gutman}. Compared to \cite{gutman} we make a new look into the compromises among error events.
We still assume that
the observations upon which the decision making is performed are
available from the source without noise. A further expansion of this
subject could restrict the decision making within
corrupted source samples.

\section{Models of source and HT}

Let ${\cal X}$ and ${\cal S}$ be finite sets: the alphabet of an
information source and its states, respectively. Let ${\cal P}({\cal
X})$ be the set of all probability distributions (PD) on ${\cal X}$.
The source in our focus is defined by the following family of
conditional PD's $G^*_s$ depending on arbitrarily and not
probabilistically varying source state $s\in {\cal S}$:
\begin{equation} {\cal G }^* \df \{G^*_s, \, s\in {\cal S}\}
\label{source-distribution}
\end{equation}
with $ G^*_s \df\{ G^*(x|s), \quad x \in {\cal X}\}. $
An output source vector ${\bf x} \df ~(x_1,..., x_N)\in $ ${\cal
X}^{N}$ will have the following probability if dictated by a state vector ${\bf s} \in
{\cal S}^{N}$:
$
G^{*N}({\bf x}|{\bf s})\df G^*_{\bf s}({\bf x})\df  \prod_{n=1}^N
G^*(x_{n}{|} s_n).
$
Furthermore, the probability of a subset ${\cal A}_N \subset{{\cal
X}^{N}}$ subject to ${\bf s} \in {\cal S}^{N}$ is measured by the
sum $ G^{*N}({\cal A}_N|{\bf s}) \df G^*_{{\bf s}}({\cal A}_N) \df
\sum\limits_{{\bf{x}}\in {\cal A}_N}G^*_{\bf s}({\bf x}). $

Our model of HT is determined by $M+1$ hypotheses about the source
distribution (\ref{source-distribution}):
$$
H_{m}:\, {\cal G}^* = {\cal G}_m, \quad  H_R: \, \mbox{none of $H_m$'s is true}
$$
with
\begin{equation}
{\cal G }_m \df \{G_{m,s}, \, s\in {\cal S}\}, \label{family-distr}
\end{equation}
where $G_{m,s} \df \{ G_m(x|s), \; x \in {\cal X}\}$, $s \in
{\cal S}$, $m=\overline{1,M}$. Let $G_m$ be the stochastic matrix defined by
(\ref{family-distr}). Based on $N$ observations of the source one
should make a decision in favor of one of those hypotheses.
Typically it can be performed by a decision maker/detector applying
a test $\varphi_{N}$ as a partition of ${\cal X}^N$ into $M+1$
disjoint subsets ${\cal A}_N^m,\; m=\overline{1,M}$ and ${\cal A}_N^R$. If
${\bf x}\in {\cal A}_N^m$ then the test adopts the hypothesis $H_m$.
If ${\bf x}\in {\cal A}_N^R$, the test rejects all the hypotheses
$H_m$, $m=\overline{1,M}$. The test design aims at achieving certain
levels of errors during the process of decision making. $(M+1)M$
different kinds of errors, denoted by $\alpha_{l,m}(\varphi_{N})$
and $\alpha_{R,m}(\varphi_{N})$, $l \neq m=\overline{1,M}$, are
possible. The probability of an erroneous acceptance of the
hypothesis $H_{l}$ when $H_{m}$ was true is
\begin{equation}
 \alpha_{l,m}(\varphi_N)\df \max_{{\bf s}\in {\cal
S}^{N}}{G_{m}^N({\cal A}_{N}^{l}|{\bf s})},\quad 1\le l\neq m \le M.
\label{errorProbAVS}
\end{equation}
And the error probability of false rejection when $H_m$ was true is
defined by
\begin{equation}
 \alpha_{R,m}(\varphi_N)\df \max_{{\bf s}\in {\cal
S}^{N}}{G_{m}^N({\cal A}_{N}^{R}|{\bf s})},\quad m = \overline{1,
M}. \label{errorProbAVS_R}
\end{equation}
Another type of error can be observed related to wrong decision in
case of true $H_m$ with the probability
\begin{eqnarray}
 \alpha_{m}(\varphi_N)& \df & \max_{{\bf s}\in {\cal
S}^{N}}{G_{m}^N(\overline{{\cal A}_{N}^{m}}|{\bf s})} \nonumber \\
&= &\sum_{l\ne m}^M{\alpha_{l,m}(\varphi_N)+ \alpha_{R,m}(\varphi
_{N})},\, m = \overline{1,M}. \label{errorProbAVS_mm}
\end{eqnarray}
So we study the following error probability
exponents/relia\-bi\-li\-ties ($\log$-s and $\exp$-s being to the
base $2$) by (\ref{errorProbAVS}) and (\ref{errorProbAVS_R}):
\begin{equation}
E_{l|m}(\varphi ){\df}\limsup_{N \rightarrow \infty
}{-\frac{1}{N}\log \alpha_{l|m}^N(\varphi_N)}, \quad l\neq
m=\overline{1,M}, \label{errorexpProbAVS_lm}
\end{equation}
\begin{equation}
E_{R,m}(\varphi ){\df}\limsup_{N \rightarrow \infty
}{-\frac{1}{N}\log \alpha_{R,m}^N(\varphi_N)}, \quad
m=\overline{1,M}, \label{errorexpProbAVS_Rm}
\end{equation}
where $\varphi \df \{\varphi_N\}_{N=1}^\infty$. From
(\ref{errorProbAVS_mm}) and (\ref{errorexpProbAVS_lm}) it follows
that
\begin{equation}
E_{m}(\varphi)=\min_{l\ne m}\left[E_{l|m}(\varphi ),
E_{R,m}(\varphi)\right]. \label{errorexpProbAVS_mm}
\end{equation}

In view of achievability concept \cite{T05} for reliabilities in
$M$-ary HT, consider the $M(M+1)$-dimensional point  ${\bf E} \df
\{E_{R,m},E_m\}_{m=\overline{1,M}}$ with respect to the error
exponents pairs $(-\frac {1}{N}\log \alpha_{R,m}(\varphi_N), -\frac
{1}{N}\log \alpha_{m}(\varphi_N))$, where the decision regions
${\cal A}_N^m$ ($m=\overline{1,M}$) and ${\cal A}_N^R$ satisfy
${\cal A}_N^m\cap {\cal A}_N^l = \emptyset $ for $m\neq l$, ${\cal
A}_N^m\cap {\cal A}_N^R = \emptyset $ and $\bigcup\limits_m{\cal
A}_N^m = {\cal X}^N/{\cal A}_N^R$.

\noindent {\bf Definition 1.}
${\bf E}$ is called achievable if for all
$\varepsilon > 0$ there exists a decision scheme $\{{\cal
A}_N^m\}_{m=1}^M$ and ${\cal A}_N^R $ with the properties
$$
-\frac {1}{N}\log\alpha_{R,m}(\varphi_N) > E_{R,m} - \varepsilon, \; -\frac {1}{N}\log\alpha_{m}(\varphi_N) > E_{m} - \varepsilon
$$
for $N$ large enough. Let ${\cal R}_{\mbox{\scriptsize AVS}}(M,R)$
denotes the set of all achievable reliabilities.

\section{Basic Properties}
Here we resume some necessary material on the typical sequences
\cite{Tom_Cov}. Let ${\cal P}({\cal S})\df \{P(s), \, s \in {\cal
S}\}$ be the collection of all PD's on ${\cal S}$ and let $ P G$ be
a marginal PD on ${\cal X}$ defined by $ P G(x) \df
\sum\limits_{s\in {\cal S}}P(s)G(x|s),\, x\in{\cal X}$.

The type of the vector ${\bf s} \in {\cal S}^{N}$ is the empirical
PD $ P_{\bf s}(s) \df \frac 1 {N}N(s|{\bf s})$, where $N(s|{\bf s})$
is the number of occurrences of $s$ in ${\bf s}.$ Let's denote the
set of all types of $N$-length state vectors by ${\cal P}^N({{\cal
S}})$. For a pair of sequences ${\bf x}\in {\cal X}^{N}$ and ${\bf
s}\in {\cal S}^{N} $ let $N(x,s|{\bf x},{\bf s})$ be the number of
occurrences of $(x,s)$ in $\{x_n,s_n\}_{n=1}^N$. The conditional
type $G_{{\bf x},{\bf s}}$ of the
 vector ${\bf x}$ with respect to the vector ${\bf s}$ is
defined by
\begin{equation}
G_{\bf x,s}(x|s) \df N(x,s|{\bf x},{{\bf s}})/N(s|{{\bf s}}), \quad
x\in {\cal X}, \, s \in {\cal S}. \label{condType}
\end{equation}

The joint type of vectors ${\bf x}$ and ${\bf s}$ is the PD $
P_{{\bf s}}\circ G_{{\bf x},{\bf s}} \df \{P_{{\bf s}}(s)G_{{\bf
x},{\bf s}}(x|s), \; x \in{\cal X},\, s\in {\cal S}\}$. For brevity
the type notations can be used without indices. Let ${\cal
G}^{N}({\cal X}|{\cal S})$ be the set of all conditional types
(\ref{condType}) and ${\cal G}({\cal X})$ be the set of all
distributions defined on $ {\cal X}$. Denote by ${\cal
T}^N_{G}(X|{\bf s})$ the set of vectors ${\bf x}$ which have the
conditional type $G$ for given ${\bf s}$ having type $P$. Let the
conditional entropy of $G$ given type  $P$ be $H(G|P)$.
The notation $H(Q)$ will stand for the unconditional entropy of
$Q\in {\cal P}({\cal X})$. Denote by ${D( G
\parallel G_{m}|P)}$ the KL
 divergence between $G $ and $G_{m}$ given type $P$ and
by $D(P G\parallel  P G_{m})$ the one between marginals $PG$ and $P G_{m}$.
The following inequality holds for every $G_{m} \in {\cal G}_m$:
\begin{equation}\label{Property1}
D(G\parallel  G_{m}|P) \ge D( PG\parallel PG_{m}).
\end{equation}
We need the next properties:
\begin{equation} \label{Property2}
|{\cal G}^{N}({\cal X}|{\cal S})|< (N+1)^{|{\cal X}| |{\cal S}|},
\end{equation}
\begin{equation}
\label{Property3} |{\cal T}^N_{G}(X|{\bf s})|\leq \exp\{NH(G|P)\}.
\end{equation}
For a PD $G_{m}\in {\cal G}({\cal X}|{\cal S})$ the sequence ${\bf
x}\in {\cal T}^N_{G}(X|{\bf s})$ has the probability
\begin{equation}
\label{Property4}
 G_{m}^N({\bf x}|{\bf s}) = \exp\{-N[ H(G|P)+ D(G\parallel G_{m}|P)]\}.
\end{equation}
(\ref{Property3}) and (\ref{Property4}) give an estimate for
conditional type class probability
\begin{equation}
\label{Property5.a} G^N_{m}({\cal T}^N_{G}(X|{\bf s})|{\bf s}) \ge
(N+1)^{|{\cal X}||{\cal S}|} \exp\{-ND(G
\parallel G_{m}|P)\},
\end{equation}
\begin{equation}
 G^N_{m}({\cal T}^N_{G}(X|{\bf s})|{\bf s})
  \le  \exp\{-ND(G\parallel G_{m}|P)\}.
\label{Property5.b}
\end{equation}

\section{Region of Acheivable Reliabilities}

Introduce the following convex hulls for each $m=\overline{1,M}$
\begin{equation}
\label{ConvexHull}
 {\cal W}_m \df \{W_m(x) \df  \sum\limits_{s\in{\cal
S}}{\lambda_s}{G}_{m,s}(x|s) \},
\end{equation}
where $x\in {\cal X},\,  0\le\lambda_s\le 1,  \,  \sum\limits_{s\in
{\cal S}}\lambda_s = 1 $,
and the region
\begin{eqnarray}
{\cal E}_{\mbox{\scriptsize AVS}}(M,R)& \df \quad  \{ {\bf E}: \quad
\forall \, W \; \, \exists \; m \; (m = \overline{1,M}),  \,  \mbox{
   s. t.} \nonumber \\   & \min\limits_{W_m\in {\cal W}_m}\! D(W\parallel
W_m)>E_{m}  \mbox{ and } \exists \, W  \mbox{ s. t.} \nonumber \\ &
\min\limits_{W_m\in {\cal W}_m}\! D(W\parallel W_m)\!>\!E_{R,m}
\mbox{ for all } m  \}. \label{singleSet_E}
\end{eqnarray}

Our main result shows that (\ref{singleSet_E}) completely
characterizes ${\cal R}_{\mbox{\scriptsize AVS}}(M,R)$.
\begin{theorem} ${\cal E}_{\mbox{\scriptsize AVS}}(M,R)$ is an
achievable region of reliabilities  $ {\cal E}_{\mbox{\scriptsize
AVS}}(M,R) \subset {\cal R}_{\mbox{\scriptsize AVS}}(M,R)$.
Moreover, if ${\bf E}\in {\cal R}_{\mbox{\scriptsize AVS}}(M,R)$,
then for any $\delta
>0$, ${\bf E}_{\delta}\in {\cal E}_{\mbox{\scriptsize AVS}}(M,R)$, where
$ {\bf E}_{\delta} \df \{E_{R,m} - \delta, E_{m} - \delta\}_{m =
\overline{1,M} }. $\label{Theorem M_HT_AVS}
\end{theorem}

\begin{IEEEproof} For the direct part, if ${\bf E} \in ~{\cal E}_{\mbox{\scriptsize
AVS}}(M,R)$, then from (\ref{Property1}), (\ref{Property3}),
(\ref{Property4}) and (\ref{Property5.b}) for any type $G\in ~{\cal
G}^N({\cal X}|{\cal S})$  and ${\bf s}\in {\cal S}^{N}$ with type
$P_{\bf s} = P$ we have
$$
G_{m,s}^N(\overline{\cal A}_N^m | {\bf s})   =  \sum\limits_{{\bf
x}\in\overline{\cal A}_N^m}G^N_{m,s}({\bf x}|{\bf s})
$$
$$
\le  \sum\limits_{{\cal T }_{G}^N(X|{\bf s})\subset \overline{\cal
A}_N^m}{\exp\{-ND(G\parallel G_{m,s}|P)\}} \nonumber
$$
\begin{equation} \le  |{\cal
G}^N({\cal X}|{\cal S})|{\exp\{-ND(PG\parallel PG_{m,s})\}}.
\label{prob_div}
\end{equation}

 For every $W_m\in {\cal W}_m$ there exists ${\bf s}\in {\cal
S}^N$, such that $W_m = P_{\bf s}G_{m,s} $. Hence, from
(\ref{prob_div}) and (\ref{Property2}) we come to
\begin{eqnarray*}
\alpha_{m}(\varphi_N) & \le & |{\cal G}^N({\cal X}|{\cal
S})|{\exp\{-N\min\limits_{W_m}{D(W\parallel W_m)}\}} \nonumber \\ &
\nonumber \le & |{\cal G}^N({\cal X}|{\cal S})|{\exp\{-NE_{m}\}}
\\ &
 \le & {\exp\{-N(E_{m}- \delta )\}}.
\end{eqnarray*}

In the same way we could get the necessary inequality for
$\alpha_{R,m}(\varphi_N)$, that is
\begin{equation}
\alpha_{R,m}(\varphi_N)  \le  {\exp\{-N(E_{R,m}- \delta )\}}.
\end{equation}
This closes the proof of the direct part.

For the converse we assume that ${\bf E}\in {\cal
R}_{\mbox{\scriptsize AVS}}(M,R)$. This provides that for every
$\varepsilon
> 0$ there exists a decision scheme
$\{{\cal A}_N^m,{\cal A}_N^R\}_{m=1}^M$ that makes the following
inequalities true as soon as $N
> N_0(\varepsilon)$:
\begin{equation}
-\frac {1}{N}\log \alpha_{R,m}(\varphi_N) > E_{R,m} - \varepsilon, \, -\frac {1}{N}\log \alpha_{m}(\varphi_N) > E_{m} - \varepsilon,
\label{converseAssumption}
\end{equation}
for all $m$'s. Pick a $\delta
>0$ and show that
\begin{equation}
\forall \, W \; \, \exists \, m \, \mbox{ s.
t.}  \min\limits_{W_m\in {\cal W}_m} D(W\parallel W_m)>E_{m}-\delta,
\label{directAssumption1}
\end{equation}
\begin{equation}
\exists \, W \mbox{ s. t.}
  \min\limits_{W_m\in {\cal W}_m} D(W\parallel W_m)>E_{R,m}-\delta
\mbox{ for all } m.
 \label{directAssumption2}
\end{equation}

For that we prove the next fact. For every $ { W}_{m}\in { {\cal
W}}_{m} $ and ${\cal A}_N \subseteq {\cal X}^{N}$ the inequality
holds:
\begin{equation}
{W}^N_{m}({\cal A}_N) \le \max_{{\bf s} \in {\cal
S}^{N}}G^N_{m}({\cal A}_N|{\bf
s}).\label{2error&convexProbInequality}
\end{equation}
To show (\ref{2error&convexProbInequality}), first note that for $
{W}_{m} \in  {\cal W}_{m}$ there exists a collection of $\lambda_{
s}$'s (by (\ref{ConvexHull}))
s.t. $ {W}_{m} =
\sum\limits_{{s}\in {\cal S}} \lambda_{{s}}G_{m,s}. $ \noindent
Whence, for $\lambda_{{\bf s}}\df \prod\limits_{n=1}^{N}
\lambda_{s_{n}}$ and any ${\cal A}_N\in {\cal X}^{N}$,$\; {\bf x}\in
{\cal A}_N$, the following estimate implies
\begin{eqnarray}
 W_m^N({\bf x}) & = & \prod_{n=1}^N{W_m( x_n)} \nonumber \\ & =&\prod\limits_{n=1}^N\sum\limits
_{s\in {\cal S}}{\lambda_s}{G}_m(x_n|s) \nonumber \\ & = &
\sum\limits _{{\bf s}\in {\cal S}^{N}}{\lambda_{\bf
s}}\prod\limits_{n=1}^N{G}_m(x_n|s_n) \nonumber \\ & \le &
\max\limits_{{\bf s}\in {\cal S}^{N}}
\prod\limits_{n=1}^N{G}_m(x_n|s_n)  \nonumber \\ & \le &
\max\limits_{{\bf s}\in {\cal S}^{N}}G_{m}^N({\bf x}|{\bf
s}).\nonumber
\end{eqnarray}
Therefore
$$ {W}^N_{m}({\cal A}_N) \le
\max\limits_{{\bf s} \in {\cal S}^{N}}G^N_{m}({\cal A}_N|{\bf s})
$$
for every $ { W}^N_{m}\in { {\cal W}}^N_{m} $ and ${\cal
A}_N \subseteq {\cal X}^{N}$.
Turning to (\ref{directAssumption1}), by the continuity of
$D(\cdot\parallel W_m)$
there exists a type
$Q\in {\cal P}^N({\cal X})$ that for $N>N_1(\varepsilon)$ and a fixed $m$ satisfies
\begin{equation} D(Q\parallel ~W_m)\le~ D(W
\parallel~ W_m) +~ \delta/2 \label{Q_Type_W}.\end{equation}
Let $\overline{W}_{m}\df \arg\min\limits_{W_m\in {\cal
W}_m}D(Q\parallel W_m)> E_{m}- \delta/2$, then in light of
(\ref{2error&convexProbInequality}) and (\ref{Property3}) we have
\begin{eqnarray*}
\alpha_{m}(\varphi_N) & \ge & \overline{W}^N_{m}(\overline{{\cal
A}_N^m})\nonumber
\\&  \ge & \overline{W}^N_{m}(\overline{{\cal A}_N^m} \cap {\cal T}_Q^N({X}) )
\nonumber \\ & \nonumber = & \sum\limits_{\overline{{\cal A}_N^m}
\cap {\cal T}_Q^N({X}) }\exp\{-N[H(Q) \nonumber \\ \nonumber&&
+D(Q\parallel \overline{W}_{m})]\}
\\ &\nonumber
\ge & |\overline{{\cal A}_N^m} \cap {\cal T}_Q^N({X}) |\exp\{-NH(Q)\} \times \nonumber \\
& &\times \exp\{-ND(Q\parallel \overline{W}_{m})\}.
\end{eqnarray*}
Note that
$
|\overline{{\cal A}_N^m} \cap {\cal T}_Q^N({X}) |\exp\{-NH(Q)\} \ge
{\exp\{-N\delta/4\}}
$
for $N > N_2(\delta)$. It follows from the inequality
$
|\overline{{\cal A}_N^m} \cap {\cal T}_Q^N({X}) | \ge \frac {|{\cal
T}_Q^N({X})|}{M}
$
which implies that
\begin{eqnarray}
|\overline{{\cal A}_N^m} & \cap & {\cal T}_Q^N({X}) |\exp\{-NH(Q)\}
\nonumber
\\ \nonumber   & \ge &  |{\cal T}_Q^N({X})|\exp\{-NH(Q)\}
\exp\{-N\frac{\log M}{N}\} \\
 & \ge &{\exp\{-N\delta/4\}}. \label{capacity}
\end{eqnarray}
 Whence, for $N > \max\{N_1(\delta),N_2(\delta)\}$ we have
\begin{eqnarray}
\alpha_{m}(\varphi_N) & \ge &  \exp\{-N[D(Q\parallel
\overline{W}_{m}) - \delta/4]\} \nonumber \\ & \nonumber \ge &
\exp\{-N[D(W\parallel \overline{W}_{m}) + \delta/4]\}
\end{eqnarray}
that with (\ref{converseAssumption}) and  $\varepsilon = 3\delta /4$
gives
$$
E_{m} - \delta < -\frac {1}{N}\log{\alpha_{m}(\varphi_N)} <
D(W\parallel \overline{W}_{m})
$$
for $N > \max\{N_0(\varepsilon),N_1(\delta),N_2(\delta)\}$ and for
every $m=\overline{1,M}$.

Now we have to proceed with the proof of (\ref{directAssumption2}).
Suppose again $\overline{W}_{m}\df \arg\min\limits_{W_m\in {\cal
W}_m}D(Q\parallel W_m)> E_{m}- \delta/2$. For a picked $\delta >0$,
if ${\bf E}_{\delta} \notin {\cal E}_{\mbox{\scriptsize AVS}}(M,R)$
then
$ \forall W \;\exists\; m \mbox{  satisfying  }
D(W \parallel \overline{W}_m)\le E_{R,m} - \delta.
$

According to (\ref{2error&convexProbInequality}), (\ref{Property3}),
(\ref{Q_Type_W}) and (\ref{capacity}) we have
\begin{eqnarray*}
\alpha_{R,m}(\varphi_N) & \ge & \overline{W}^N_{m}({{\cal
A}_N^R})\nonumber
\\&  \ge & \overline{W}^N_{m}({{\cal A}_N^R} \cap {\cal T}_Q^N({X}) )
\nonumber \\ & \nonumber = & \sum\limits_{{{\cal A}_N^R} \cap {\cal
T}_Q^N({X}) }\exp\{-N[H(Q) \nonumber \\ \nonumber&& +D(Q\parallel
\overline{W}_{m})]\} \nonumber \\
& \ge & |{{\cal A}_N^R} \cap {\cal T}_Q^N({X}) |\exp\{-NH(Q)\} \times \nonumber \\
& &\times \exp\{-ND(Q\parallel \overline{W}_{m})\} \\ \nonumber &
\ge & \exp\{-N[D(W\parallel \overline{W}_{m}) - \delta/4]\} \\
\nonumber & \ge & \exp\{-N[E_{R,m} - \delta/4]\}.
\end{eqnarray*}

However the last inequality is in conflict with
(\ref{converseAssumption}) for $\varepsilon < \delta / 4$ and $N >
\max\{N_0(\varepsilon),N_1(\delta),N_2(\delta)\}$.
\end{IEEEproof}

\section{Optimal decision schemes}

Here we look for optimal decision schemes and the corresponding best
error exponents in the following sense (similar to LAO test
\cite{H90}, \cite{AH06}). Let $E_m, \; m= \overline{1,M},$ be fixed:
what are the ``maximum" values for $\{E^*_{l,m}, E^*_{R,m}\}_{l\neq
m=\overline{1,M}}$ such that there is no other $\{E^{\prime}_{l,m},
E^{\prime}_{R,m}\}_{l\neq m=\overline{1,M}}$ satisfying
$E^{\prime}_{l,m}
> E^{*}_{l,m} \mbox{ and  } E^{\prime}_{R,m}
> E^*_{R,m}$ for all $l\neq m = \overline{1,M}$? Con\-si\-der the
following test sequence $\varphi^{*}$ in terms of the sets
$$
{\cal B}_R \df \{W:\, \min_{W_m \in {\cal W}_m}D(W \parallel W_m )
> E_m \mbox{ for all } m\},
$$
$$
{\cal B}_m \df \{W:\, \min_{W_m \in {\cal W}_m}D(W \parallel W_m ) <
E_m\}, \quad m=\overline{1,M}.
$$
Define ($l \neq m= \overline{1,M}$):
\begin{equation}
E_{R,m}(\varphi^{*})\df E_{R,m}^{*} \df \min_{W \in {\cal
B}_R}\min_{W_m \in {\cal W}_m}{D(W\parallel W_m)},
\label{optimalErrorExponentRm}
\end{equation}
\begin{equation}
E_{l,m}(\varphi^{*}) \df E_{l,m}^{*} \df \min_{W \in {\cal
B}_l}\min_{W_m \in {\cal W}_m}{D(W\parallel W_m)}.
\label{optimalErrorExponentlm}
\end{equation}
\begin{theorem}
\label{AVS_optimality} Let the following inequalities hold:
$$
E_{1}^{*} < \min\limits_{m}\{\min_{W_m\in {\cal W}_m,W_1\in {\cal
W}_1}D(W_m
\parallel W_1)\},
$$
$$
E_{m}^{*} < \min\limits_{l \neq
m}\{\min_{l=\overline{1,m-1}}E_{l,m},\min_{l=\overline{m+1,M}}\min_{{W_l
\in {\cal W}_l,} \atop{W_m \in {\cal W}_m}}D(W_l
\parallel W_m)\},
$$
then there exist optimal sequence of tests and the corresponding
optimal vector of reliabilities are defined as in
(\ref{optimalErrorExponentRm})-(\ref{optimalErrorExponentlm}).
\end{theorem}

\begin{IEEEproof}
Let the decision on $R$ or an $m$ be made based on the partition:
$
{\cal D}_m \df \bigcup_{W\in {\cal B}_m}{{\cal T}_W^N(X)},
$
$
{\cal D}_R \df \bigcup_{W\in {\cal B}_R}{\cal T}_W^N(X).
$
Note that ${\cal D}_m\cap {\cal D}_l \neq \emptyset$ and ${\cal
D}_m\cap {\cal D}_R \neq \emptyset$, $m\neq l=\overline{1,M}$.

For $\overline{W}_m \df arg\min\limits_{W_m\in {\cal
W}_m}D(W\parallel W_m)$, $m=\overline{1,M}$,  and $\varphi \df
\{\varphi^*_N\}_{N=1}^\infty$ perform (applying unconditional verion
of (\ref{Property5.a}))
\begin{eqnarray*}
\alpha_{R,m}(\varphi_N) & \ge & \overline{W}^N_{m}({\cal
D}_{R})\nonumber
\\ &  \ge & \overline{W}^N_{m}(\bigcup_{W\in {\cal B}_R} {\cal T}_W^N({X}) ) \nonumber \\ &\nonumber
\ge & \max_{W\in {\cal B}_R} \exp\{-N[D(W\parallel \overline{W}_{m})
+ o_N(1)]\}
\\ & = & \exp\{-N[\min_{W\in {\cal B}_R}D(W\parallel \overline{W}_{m})
+ o_N(1)]\}.
\end{eqnarray*}
In a similar way we can obtain the inequality
\begin{equation}
\alpha_{l,m}(\varphi_N)  \ge  \exp\{-N[\min\limits_{W_m\in {\cal
W}_m} \! \min\limits_{W\in {\cal D}_l}D(W\parallel W_{m})  +
o_N(1)]\}. \label{optimalInequality_lm}
\end{equation}
The proof of the converse inequalities
\begin{equation}\alpha_{R,m}(\varphi_N) \le \exp\{-N[\min\limits_{W_m \! \in\! {\cal
W}_m}\!\min\limits_{W\in {\cal
 D}_R}\!D(W\parallel W_{m}) + o_N(1)]\}
\label{converseOptimalInequality_Rm}\end{equation}
\begin{equation}\alpha_{l,m}(\varphi_N)  \le  \exp\{-N[\min\limits_{W_m\in {\cal
W}_m}\min\limits_{W\in {\cal D}_l}D(W\parallel W_{m}) + o_N(1)]\}
\label{converseOptimalInequality_lm}\end{equation} are omitted here
because of space restrictions.

Taking into account (\ref{optimalInequality_lm}),
(\ref{converseOptimalInequality_Rm}),
(\ref{converseOptimalInequality_lm}) and the continuity of the
functional $D(W \parallel W_m)$ we obtain that the limit
$\lim\limits_{N \rightarrow \infty}{\{\sup
{-N^{-1}\log{\alpha_{l,m}^N(\varphi_N^* )\}}}}$ exists and equals to
$E_{l,m}^*$.

The proof will be accomplished if we demonstrate that $\varphi^*$ is
optimal. Let $\varphi^{\prime}$ be a test defined by the sets
$({\cal D}^{\prime}_{m},{\cal D}^{\prime}_{R})$ s.t.
$$
E^{\prime}_{l,m} > E_{l|m}^*, \; E^{\prime}_{R,m} > E_{R|m}^*, \; l \neq m = \overline{1,M}.
$$
It yields for $N$ large enough that
$$
\alpha_{l,m}^{N}(\varphi_N^{\prime})<
\alpha_{l,m}^{N}(\varphi_N^{*}), \quad \alpha_{R,m}^{N}(\varphi_N^{\prime})<
\alpha_{R,m}^{N}(\varphi_N^{*}).
$$
Below we examine the relation between $ ({\cal D}_{m}, {\cal D}_{R})$
and $ ({\cal D}^{\prime}_{m}, {\cal D}^{\prime}_{R})$. Four cases
are possible:

1) ${\cal D}_{m} \cap {\cal D}_{m}^{\prime} = \emptyset  $,

2) ${\cal D}_{m} \subset {\cal D}_{m}^{\prime} $,

3) ${\cal D}^{\prime}_{m} \subset {\cal D}_{m} $,

4) ${\cal D}_{m} \cap {\cal D}_{m}^{\prime} \neq \emptyset $.

\noindent The same cases exist also for ${\cal D}_{R}$ and ${\cal
D}_{R}^{\prime}$.

Consider ${\cal D}_{m} \cap {\cal D}_{m}^{\prime} = \emptyset $
case. It follows that there exists $l \neq m$ such that ${\cal
D}_{m} \cap {\cal D}_{l}^{\prime} \neq \emptyset $. That is $\exists
W$ such that $D(W \parallel \overline{W}_m ) < E_m^*$, so ${\cal
T}_W^N({X}) \subset {\cal D}_{l}^{\prime}$. Compute
\begin{eqnarray*}
\alpha_{l,m}^{N}(\varphi_N^{\prime}) & = & \max\limits_{{\bf s} \in
{\cal S}^N} G_{m} ({\cal D}_{l}^{\prime}|{\bf s}) \nonumber
\\ &  \ge & \overline{W}^N_{m}( {\cal T}_W^N({X}) ) \nonumber \\ &\nonumber
\ge &  \exp\{-N[D(W\parallel \overline{W}_{m})  + o_N(1)]\}
\\ &\nonumber = & \exp\{-N[D(W\parallel \overline{W}_{m})
+ o_N(1)]\} \\ &\nonumber = & \exp\{-N[E^*_m + o_N(1)]\}.
\end{eqnarray*}
Thus $E_{l,m}^{\prime} < E_m^{\prime}= E_m^*$ which contradicts to
(\ref{errorexpProbAVS_mm}).
\end{IEEEproof}

\section{Geometric interpretations}
Fig. 1 presents a geometric interpretation for the decision scheme
in Theorem \ref{Theorem M_HT_AVS}. Relevantly, Fig. 2 and 3 illustrate
the geometry of the Chernoff bounds derived in \cite{GH10} for the multi-HT where the rejection
is not an alternative (c.f. \cite{Brandon} for DMS's). Those interpretations are comprehensible with conceptual details given in \cite{GH10}.

\section{Results for DMS}

With assumption of ${\cal S} = 1$ we get the model of multi-HT with
rejection for DMS:
$$
H_{m}:\, {G}^* = {G}_m, \;  H_R: \,\mbox{none of $H_m$'s is true},
$$
with $ {G }_m \df \{G_{m}(x), \, x\in {\cal X}\}$, $m=\overline{1,M}$. The problem here
is to make a decision regarding the generic ${G}^*$ among $M$
alternative PD's $G_m,\; m=\overline{1,M}$, and the rejection. Let
\begin{eqnarray}
{\cal E}(M,R) \df &  \{ {\bf E}: \quad \forall \, Q \; \, \exists \;
m \; (m = \overline{1,M}),  \,  \mbox{
   s. t.} \nonumber \\  &  D(Q\parallel
G_m)>E_{m}  \mbox{ and } \exists \, Q  \mbox{ s. t.} \nonumber \\
&   D(Q\parallel G_m)>E_{R,m} \mbox{ for all } m  \}.
\nonumber
\end{eqnarray}
\begin{theorem} Theorem  \ref{Theorem M_HT_AVS} implies that $ {\cal E}(M,R) \subset {\cal
R}(M,R)$. Conversely, if ${\bf E}\in {\cal R}(M,R)$, then for any
$\delta
>0$, ${\bf E}_{\delta}\in {\cal E}(M,R)$, where
$ {\bf E}_{\delta} \df \{E_{R,m} - \delta, E_{m} - \delta\}_{m =
\overline{1,M} }. $\label{Theorem M_HT_DMS}
\end{theorem}

To formulate the DMS counterpart of Theorem \ref{AVS_optimality}
define the sets:
$$
{\cal B}_R({\mbox{DMS}}) \df \{Q:\, D(Q \parallel G_m)
> E_m, \, \mbox{for all} \, m = \overline{1,M} \},
$$
$$
{\cal B}_m({\mbox{DMS}}) \df \{Q:\, D(Q \parallel G_m ) < E_m\},
\, m=\overline{1,M}.
$$
Furthermore
\begin{equation}
E_{R,m}^{*} \df \min_{Q \in {\cal B}_R({\mbox{\tiny
DMS}})}{D(Q\parallel G_m)}, \quad  m= \overline{1,M},
\label{optimalErrorExponentRm_DMS}
\end{equation}
\begin{equation}
E_{l,m}^{*} \df \min_{Q \in {\cal B}_l({\mbox{\tiny
DMS}})}{D(Q\parallel G_m)}, \quad l\neq m= \overline{1,M}.
\label{optimalErrorExponentlm_DMS}
\end{equation}

\begin{center}
\includegraphics[width=3.2in]{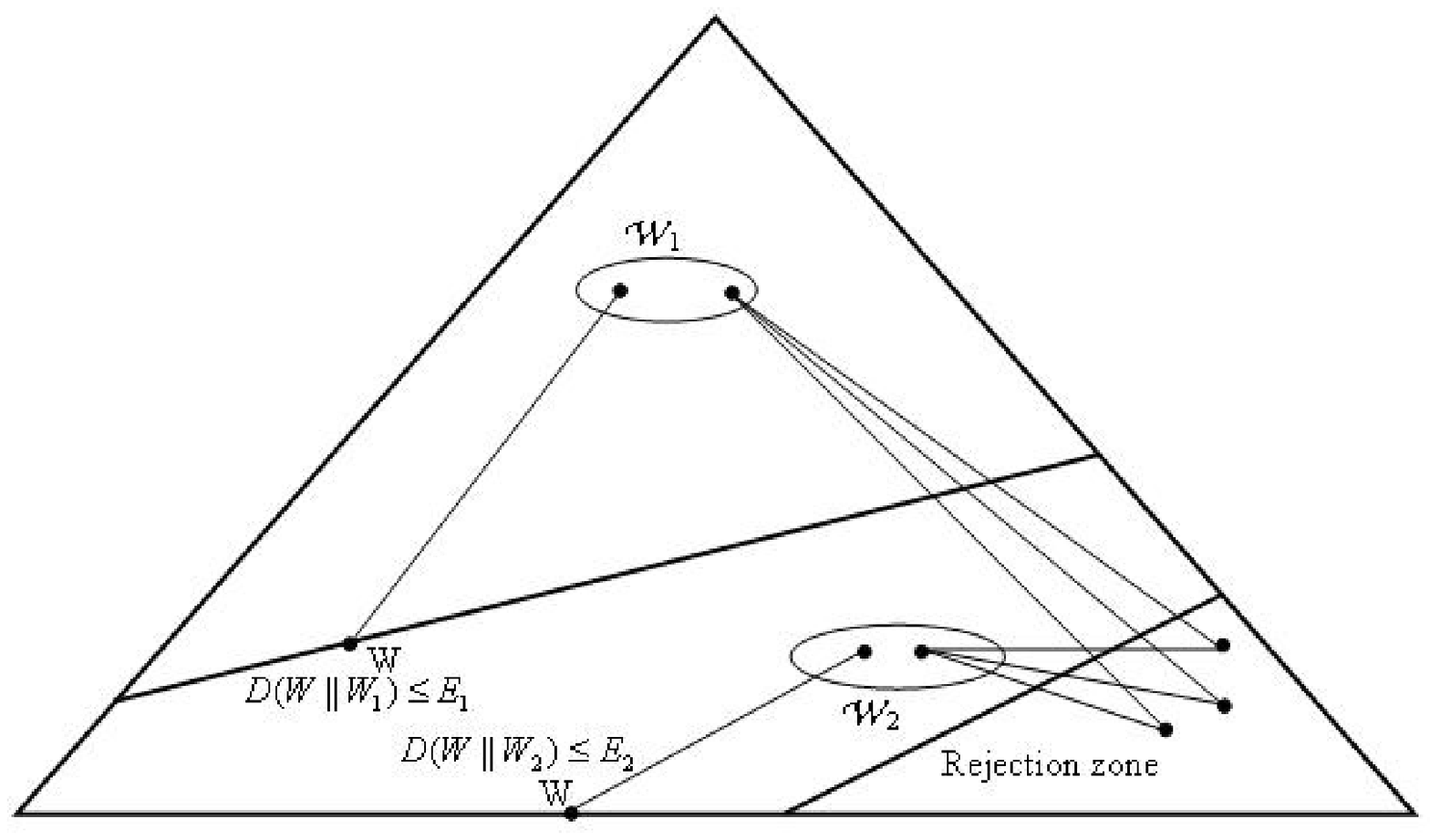}\\
     {\small Fig. 1: Multiple HT with rejection.}
\end{center}
\begin{center}
\includegraphics[width=3.2in]{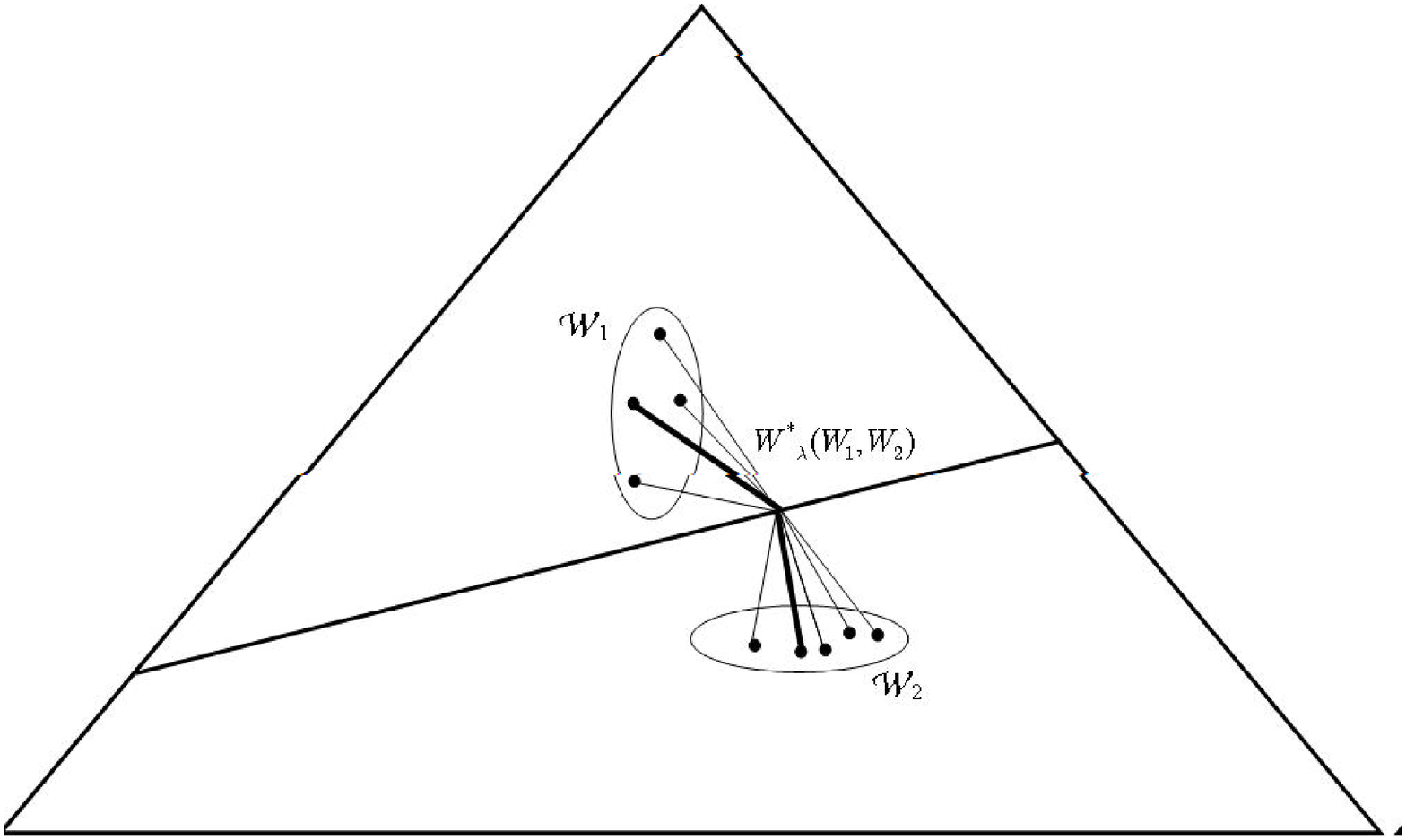}\\
     {\small Fig. 2: Chernoff bounds: binary HT: AVS.}
\end{center}
\begin{center}
\includegraphics[width=2.2in]{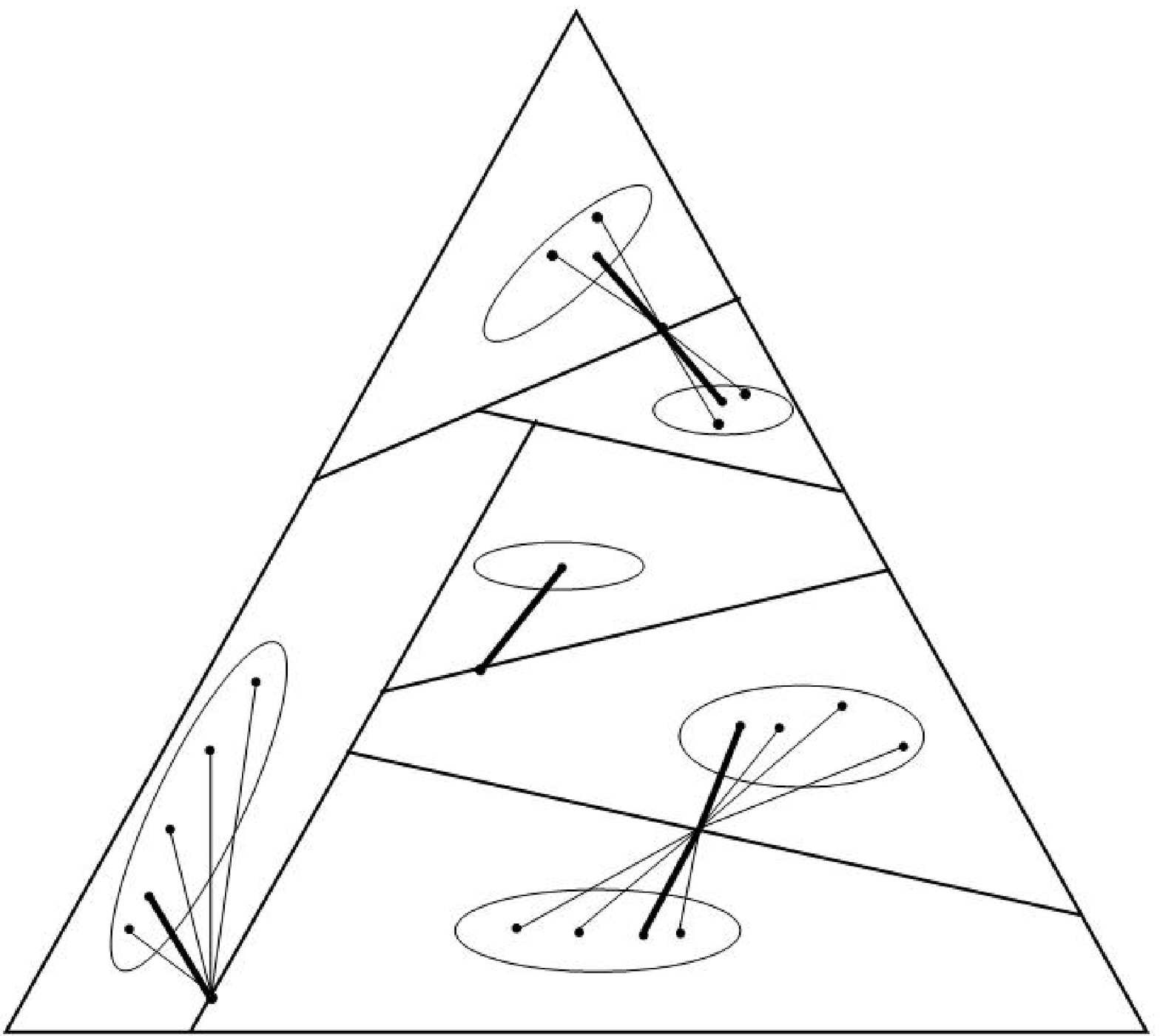}\\
     {\small Fig. 3: Chernoff bounds: multiple HT: AVS.}
\end{center}

\begin{theorem}
\label{optimalDMS}
If $D(G_m \parallel G_l) >0$, $m\neq l
=\overline{1,M}$, and
$$
E_{1}^{*} < \min\limits_{m}\{D(G_m
\parallel G_1)\},
$$
$$
E_{m}^{*} < \min\limits_{l \neq
m}\{\min_{l=\overline{1,m-1}}E_{l,m},\min_{l=\overline{m+1,M}}D(G_l
\parallel G_m)\},
$$
then there exist optimal tests and the corresponding optimal vector
of reliabilities are defined according to
(\ref{optimalErrorExponentRm_DMS})-(\ref{optimalErrorExponentlm_DMS}).
\end{theorem}

According to \cite{HarHak} the authors claim to have obtained Theorem \ref{optimalDMS} independently.
\begin{remark} It is possible to prove that
$$
 \min_{l={\overline{1,M}, \, l\neq m}}\left[E_{l,m}^{*}, E_{R,m}^{*}\right]=E_{R,m}^{*},
\quad \mbox{for all} \quad  m=\overline{1,M}.
$$
This means that {\it discrimination is always easier than rejection}.
\end{remark}

\pagebreak

\end{document}